\documentclass[sigconf]{acmart}

\AtBeginDocument{%
  \providecommand\BibTeX{{%
    \normalfont B\kern-0.5em{\scshape i\kern-0.25em b}\kern-0.8em\TeX}}}

\setcopyright{acmcopyright}
\copyrightyear{2018}
\acmYear{2018}
\acmDOI{10.1145/1122445.1122456}

\usepackage{subfigure}
\usepackage{epsfig}

\usepackage{comment}
\usepackage{multirow}
\usepackage{subfigure}
\newcommand{\model}{{HAHE}}

\newtheorem{mydef}{Definition}

\makeatletter
\newif\if@restonecol
\makeatother

\usepackage[linesnumbered,ruled,vlined]{algorithm2e}
\usepackage{algpseudocode}

\begin{document}

\title{HAHE: Hierarchical Attentive Heterogeneous Information Network Embedding}

\author{Sheng Zhou}
\email{zhousheng_zju@zju.edu.cn}
\affiliation{%
  \institution{Department of Computer Science\\ Zhejiang University}
}

\author{Jiajun Bu}
\email{bjj@zju.edu.cn}
\affiliation{%
  \institution{Department of Computer Science\\ Zhejiang University}
}

\author{Xin Wang}
\email{xin_wang@tsinghua.edu.cn}
\affiliation{%
  \institution{Department of Computer Science and Technology\\ Tsinghua University}
}

\author{Jiawei Chen}
\email{sleepyhunt@zju.edu.cn}
\affiliation{%
  \institution{Department of Computer Science\\ Zhejiang University}
}

\author{Can Wang}
\email{wcan@zju.edu.cn}
\affiliation{%
  \institution{Department of Computer Science\\ Zhejiang University}
}

\begin{abstract}
  Heterogeneous information network (HIN) embedding has recently attracted much attention due to its effectiveness in dealing with the complex heterogeneous data.  Meta path, which connects different object types with various semantic meanings, is widely used by existing HIN embedding works. However, several challenges have not been addressed so far. First, different meta paths convey different semantic meanings, while existing works assume that all nodes share same weights for meta paths and ignore the personalized preferences of different nodes on different meta paths.
Second, given a meta path, nodes in HIN are connected by path instances while existing works fail to fully explore the differences between path instances that reflect nodes' preferences in the semantic space. 
rTo tackle the above challenges, we propose a {\bf H}ierarchical {\bf A}ttentive {\bf H}eterogeneous information network {\bf E}mbedding (\model) model to capture the personalized preferences on meta paths and path instances in each semantic space. As path instances are based on a particular meta path, a hierarchical attention mechanism is naturally utilized to model the personalized preference on meta paths and path instances. 
Extensive experiments on several real-world datasets show that our proposed \model model significantly outperforms the state-of-the-art methods in terms of various data mining tasks.
\end{abstract}

\begin{CCSXML}
<ccs2012>
 <concept>
  <concept_id>10010520.10010553.10010562</concept_id>
  <concept_desc>Computer systems organization~Embedded systems</concept_desc>
  <concept_significance>500</concept_significance>
 </concept>
 <concept>
  <concept_id>10010520.10010575.10010755</concept_id>
  <concept_desc>Computer systems organization~Redundancy</concept_desc>
  <concept_significance>300</concept_significance>
 </concept>
 <concept>
  <concept_id>10010520.10010553.10010554</concept_id>
  <concept_desc>Computer systems organization~Robotics</concept_desc>
  <concept_significance>100</concept_significance>
 </concept>
 <concept>
  <concept_id>10003033.10003083.10003095</concept_id>
  <concept_desc>Networks~Network reliability</concept_desc>
  <concept_significance>100</concept_significance>
 </concept>
</ccs2012>
\end{CCSXML}

\ccsdesc[500]{Computer systems organization~Embedded systems}
\ccsdesc[300]{Computer systems organization~Redundancy}
\ccsdesc{Computer systems organization~Robotics}
\ccsdesc[100]{Networks~Network reliability}

\keywords{Heterogeneous Information Network, Embedding, Attention}
\maketitle

\section{Introduction}
A heterogeneous information network (HIN) is a network whose nodes or links share different types, many real world networks including bibliographic network, social network and knowledge base etc. can be modeled as HINs.
Meanwhile, as an efficient and effective way to represent and manage large-scale networks, network embedding maps the topological structure into low-dimensional vector space such that the original network proximity can be well preserved. The embedded results have been proved to be extremely useful as feature inputs for a wide variety of graph analysis tasks including clustering, classification and prediction \cite{hamilton2017inductive,zhang2018metagraph2vec,wang2018shine}. 
To marry the advantages of HINs with network embedding, recently, HIN embeddings are drawing much attention from both academia and industry communities.

Compared with homogeneous network, the heterogeneity in HINs brings in more information of similarity between nodes for network embedding to preserve. 
In particular, meta path, a relation sequence connecting different types of nodes, is widely used to extract structural features and capture relevance semantics between nodes in HINs. 
Although some existing HIN embedding methods \cite{fu2017hin2vec,zhang2018metagraph2vec,shang2016meta} have utilized meta paths to learn comprehensive proximity between nodes, several challenges have not been addressed so far. 

First, existing HIN embedding methods either do not distinguish meta paths or assume same weights of meta paths for all nodes. 
The personalized preferences of nodes on meta paths can not be captured and the proximity to be preserved is not complete enough. 
Take the heterogeneous social network embedding for friend recommendation as an example, some users prefer to making friends with those sharing similar tags (the meta path user-tag-user), some users prefer to making friends with those close in geography (the meta path user-location-user).
Only modeling the global preference will miss such personalization and result in inaccurate embeddings.
Although modeling the personalized preference on meta paths can help to learn better representations of nodes, it will burden users a lot to manually provide explicit guidance in determining importance of meta paths for each node. 
It is necessary to efficiently and effectively model the personal preference on the meta paths.

Second, given a meta path, nodes in HINs are connected by path instances. Most of the existing methods measure the similarity by counting the number of path instances and ignore the difference between path instances. Still take the heterogeneous social network embedding for friend recommendation as an instance, given a meta path, i.e., user-tag-user, 
a user can be connected to other users by path instances. However, the tag connects them may be rough or cult, the connected user may have many or a few tags. All the above situations will result differences between meta path instances.
Distinguishing these path instances can highlight the most relevant path instance and ignore noisy ones, as a result, better embedding results can be learned. However, existing methods fail to capture such personalized preference on path instances.

To solve the above challenges, in this paper, we propose a {\bf H}ierarchical {\bf A}ttentive {\bf H}eterogeneous information network {\bf E}mbedding (\model) model to efficiently learn HIN embedding while modeling the personalized preference on meta paths and path instances.
To meet the fact that path instances are based on given meta paths, a hierarchical attention mechanism is naturally utilized with a meta path attention layer and a path instance attention layer. The meta path attention layer learns personalized preferences towards meta paths for each individual node, the path instance attention layer determines importance of path instances with respect to meta path. 

In particular, the main advantages of using such attention in HIN embedding can be summarized as follows:
i) Attention allows the \model\ model to be robust towards noisy parts of the HINs including both meta paths and path instances, thus improve the signal-to-noise (SNR) ratio\cite{lee2018attention};
ii) Attention allows the \model\ model to assign a relevance score to each node in the HINs  to highlight nodes with the most task-relevant information, it also provides a way for us to make the model more interpretable. We will provide further discussion in the experiments.

The contributions of this paper can be summarized as follows:
\begin{enumerate}
	\item We propose a Hierarchical Attentive Heterogeneous information network Embedding (\model) model to learn HIN embeddings which captures the personalized preference on both meta paths and path instances.
	\item We elaborately design a hierarchical attention mechanism to learn attention coefficients on meta paths and path instances, we provide evidence that the learned coefficients can reflect the performance of meta path.  
	\item We conduct experiments in terms of several data mining tasks on real-world datasets to show the superiority of our model against several existing methods and give comprehensive analysis on the learned attention coefficients in order to gain more insights of the datasets.
\end{enumerate}

The remainder of this paper is organized as follows.
Section \ref{sectionrelated} introduces the related work. 
Section \ref{sectionprelit} describes notations used in this paper and presents some preliminary knowledge.
Then, we propose the \model\ model in Section \ref{sectionmodel}. 
Experiments and detailed analysis are reported in Section \ref{sectionexp}. 
Finally, we conclude the paper in Section \ref{sectionconclu}.

\section{Related Work}
\label{sectionrelated}
In this section, we will review the related studies in three aspects, namely heterogeneous information network, network embedding and heterogeneous information network embedding. 

\textbf{Heterogeneous Information Network}
As a newly emerging direction, heterogeneous information network (HIN) has been extensively studied as a powerful and effective paradigm to model complex objects and relations. 
In HIN, nodes can be reached by paths with different semantic meanings and these paths(also called meta-paths \cite{sun2011pathsim}) have been explored for fulfilling tasks, including classification \cite{ji2010graph}, clustering \cite{sun2013pathselclus,sun2012relation}, recommendation \cite{yu2014personalized,chen2017task}, and outlier detection \cite{gupta2013community}.
However, with the development of data collecting, the scale of HIN are growing and the adjacent representation is always high-dimensional and sparse.
A low-dimensional and dense representation is needed to serve as the basis for different downstream applications.

\textbf{Network Embedding}
Network embedding aims at learning low-dimensional vector representation to facilitate a better understanding of semantic relationships among nodes.
Among them, a branch of methods \cite{grover2016node2vec,perozzi2014deepwalk,tang2015line} employ a truncated random walk to generate node sequences, which is treated as sentences in language models and fed to the skip-gram model to learn embeddings. 
Beyond skip-gram model, graph structure is also incorporated into deep auto-encoder to preserve the highly non-linear first order and second order proximities \cite{tang2015line,wang2016structural,cao2015grarep}. 
Recently, inspired by GCN \cite{kipf2016semi} which use convolution operators that offer promise as an embedding  methodology, a wide variety of graph neural network(GNN) models have been proposed \cite{hamilton2017inductive,velickovic2017graph,you2018graphrnn,you2018graphrnn}.

\textbf{Heterogeneous Information Network Embedding}
Recently, network embedding has been extended to HINs and a bunch of methods have been proposed.
Among them, heterogeneous skip-gram model based methods are proposed \cite{dong2017metapath2vec,zhang2018metagraph2vec,huang2017heterogeneous} where meta-path based random walks are used to generate graph contexts.
Beyonds the meta-path based random walk and skip-gram model, neural networks are also explored on HINs.
In HIN2vec \cite{fu2017hin2vec}, a single-hidden-layer feedforward neural network is applied to enable users to capture rich semantics of relationships and the details of the network structure to learn representations of nodes in HINs. 
In HNE \cite{chang2015heterogeneous} and PTE \cite{tang2015pte}, node embeddings are learned by capturing 1-hop neighborhood relationships between nodes via deep architectures.
In Aspem \cite{shi2018aspem}, HIN is decomposed into multiple aspects and embeddings are derived from the aspects.
Some methods learn embedding on specific HINs such as knowledge graph (KG)\cite{shi2017proje}, signed HINs \cite{wang2018shine} or employ HIN embedding for specific tasks such as similarity search \cite{shang2016meta} and recommendation \cite{shi2018heterogeneous,Hu:2018:LMB:3219819.3219965,chen2017task}

\section{PRELIMINARIES}
\label{sectionprelit}
In this section, we introduce some background definitions and formally define the problem of Heterogeneous Information Network Embedding.

\begin{mydef}
  \textbf{Heterogeneous Information Network}
  A heterogeneous information network(HIN)  \cite{sun2011pathsim} is defined as a network with multiple types of nodes and/or links.
  As a mathematical abstraction, we define a HIN as $G=\{\mathcal{V},\mathcal{E}\}$, where $\mathcal{V}$ denotes the set of nodes and $\mathcal{E}$ denotes the set of links.
  A HIN is also associated with a node type mapping function $f_{v}:\mathcal{V} \rightarrow \mathcal{O}$, which maps the node to a predefined node type, and a link type mapping function $f_{e}:\mathcal{E} \rightarrow \mathcal{R}$, which maps the link to a predefined link type.
\end{mydef}

\begin{mydef}
\textbf{Network schema and meta path} Network schema is a template for a HIN ${G}$ which is a directed graph defined over object types, denoted as $T_{G}=\{\mathcal{O}, \mathcal{L}\}$.
  A meta path \cite{sun2011pathsim} $\pi$ is defined on the network schema $T_{G}=\{\mathcal{O}, \mathcal{L}\}$ and is denoted in the form of $\pi=o_{1} \stackrel{l_{1}}{\longrightarrow} o_{2} \stackrel{l_{2}}{\longrightarrow} ... \stackrel{l_{m-1}}{\longrightarrow} o_{m}$.
  A path instance $\mathcal{I}$, which goes through nodes $v_{1},v_{2},...,v_{m}$, is an instance of the meta path $\pi$, if\ $\forall i=1,...,m, o_{i}=f_{v}(v_{i})$ and $l_{i}=f_{e}(v_{i},v_{i+1})$.
  \end{mydef}

The definition of meta path is often given by users with prior knowledge. 
Some previous works \cite{sun2013pathselclus,yu2012user} focused on automatically find the meta path.
However, it is not the main problem we aim to solve in this paper and we assume the meta path is defined by user.
Given the meta-path set, our work focuses on distinguishing these meta-paths.
In real-world HIN, not all the node types are studied and we separate them into target node type and content node type.

\begin{mydef}
	\textbf{Target/Content type nodes}
	The target type nodes $\mathcal{V}_{T}$ are defined as nodes aims to be embedded in network embedding. They are often associated with labels in semi-supervised learning tasks. 
	The content type nodes $\mathcal{V}_{C}$ are defined as the rest type of nodes which serves as connection between target type nodes. 
\end{mydef}

It is worth noting that providing labels for all node types is time-consuming and labor-intensive in real-world applications. 
The target type nodes can be connected with different types of content nodes with semantic meanings. 
In this paper, we only learn the embedding for one particular target type of nodes in HIN. 
Learning embeddings for all node types can be achieved by setting each node type as target type.

\begin{mydef}
  \textbf{Heterogeneous Information Network Embedding}
  Given a heterogeneous information network(HIN) $G=\{\mathcal{V},\mathcal{E}\}$, corresponding node type mapping function $f_{v}:\mathcal{V} \rightarrow \mathcal{O}$ and edge type mapping function $f_{e}:\mathcal{E} \rightarrow \mathcal{R}$. Heterogeneous Information Network Embedding aims at learning a function $f: \mathcal{V} \rightarrow R^{d}$ that projects node $v\in \mathcal{V}_{T}$ into a vector in a d-dimensional space $R^{d}$, where $d \ll |\mathcal{V}|$.
\end{mydef}

\section{Model}
\label{sectionmodel}
\begin{figure}[h]
\centering
\includegraphics[width=0.48\textwidth]{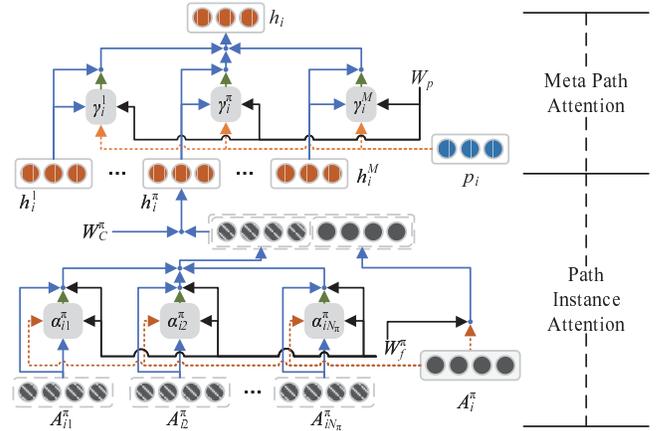}
\caption{The overall architecture of \model\ model.}
\label{model}
\end{figure}

In this section, we describe the details of our proposed Hierarchical Attentive Heterogeneous Information Network Embedding(\model) model. The architecture of \model\ is showed in Fig \ref{model}, we will introduce it from the bottom path instance attention layer to the top meta path attention layer.

\subsection{Path instance attention layer}
Given HIN $G=\{\mathcal{V},\mathcal{E}\}$ and meta path $\pi$, nodes are connected by path instances, the path instance attention layer aims at learning meta path $\pi$ based embedding $H^{\pi}$ while distinguishing the path instances.
The basic idea is to learn embedding by collecting information from node itself and nodes connected by path instances,  the path instances are distinguished based on the similarity of structural features.

We use the normalized meta path based adjacent vector $A^{\pi}_{i}$ as the structural feature representation of node $v_{i}$.
Leveraging the adjacent vector enjoys the following properties: first, nodes in different semantic space will be dowered with different structural features which can better unleash the power of meta path, second, both meta path based first and second order proximity can be preserved in meta path based embedding $H^{\pi}$.
Considering the fact that meta path based adjacent vector can be high-dimensional and sparse, we first use meta path specific feature transformation with Multilayer Perceptron(MLP), parameterized by $\mathbf{W}^{\pi}_{f}\in R^{N\times d}$ to transform the meta path $\pi$ based structural feature into $d$ dimensional space.

For nodes connects to node $v_{i}$ by path instances, those who share similar structure features with node $v_{i}$ will be assigned with large attention coefficient $\alpha^{\pi}$ which can be written as:
\begin{equation}
	s^{\pi}_{ij} = \frac{(\mathbf{W}^{\pi}_{f}A^{\pi}_{i})^{T}\cdot \mathbf{W}^{\pi}_{f}A^{\pi}_{j}}{||\mathbf{W}^{\pi}_{f}A^{\pi}_{i}||\cdot ||\mathbf{W}^{\pi}_{f}A^{\pi}_{j}||}\quad 
	\alpha^{\pi}_{ij}  = \frac{exp(s^{\pi}_{ij})}{\sum_{k\in \mathcal{N}^{\pi}_{i}}(s^{\pi}_{ik})}
\label{nal1}
\end{equation}
where $s^{\pi}_{ij}$ denotes the transformed feature similarity between node $v_{i}$ and $v_{j}$ based on meta path $\pi$, 
$\mathbf{W}^{\pi}_{f}$ denotes the structural feature transformation matrix for meta path $\pi$. $\alpha^{\pi}_{ij}$ is the path instance attention coefficients of node $v_{j}$ in learning meta path based embedding $h^{\pi}_{i}$, $\mathcal{N}^{\pi}_{i}$ is the neighborhood nodes that connect to node $v_{i}$ based on meta path $\pi$. 

Some existing work \cite{velickovic2017graph} learn embedding by combining the transformed feature of node itself with the aggregated set of nodes. However, for meta path like paper-conference-paper in bibliographic network, there may exist large amount of nodes connected by path instances, it is inefficient to aggregate information from all these nodes and node's own feature may be diluted in the aggregated features. 
To solve the above problem, we first uniformly sampled some nodes from the connected node set and learn the aggregated embedding $h^{\pi}_{N(i)}$:
\begin{equation}
	h^{\pi}_{\mathcal{N}(i)}=\sigma(\sum_{j\in \mathcal{N}^{\pi}_{i}}\alpha^{\pi}_{ij}\mathbf{W}^{\pi}_{f}A^{\pi}_{j})
\label{nal2}
\end{equation}
where $\sigma()$ is the activation function and we use Tanh function here, $h^{\pi}_{\mathcal{N}(i)}$ is aggregated embedding, $\mathcal{N}^{\pi}_{i}$ is the set of nodes connect to node $v_{i}$ with path instances. Then we concatenate node's own feature with aggregated embedding and get the meta path based embedding $h^{\pi}_{i}$:
\begin{equation}
	h^{\pi}_{i} = W^{\pi}_{C} \left [h^{\pi}_{\mathcal{N}(i)};\ \mathbf{W}^{\pi}_{f}A^{\pi}_{i} \right ]
\label{nal3}
\end{equation}
where $\mathbf{W}^{\pi}_{C}\in R^{2d\times d}$ is the weight of linear transformation from the concatenation to the embedding space, $h^{\pi}_{\mathcal{N}(i)}$ is the meta path based neighborhood embedding, $\mathbf{W}^{\pi}_{f}x^{\pi}_{i}$ is the transformed node feature, $\left [ \cdot \right ]$ is the concatenation of vectors.
To summarize, the learned meta path based embedding $h^{\pi}_{i}$ can not only keep its own feature, but also weighted combine information from nodes connected by path instances.

\subsection{Meta path attention layer}
Given meta path based embedding $\{H^{1},H^{2}...H^{M}\}\in R^{N\times d}$ learned from path instance attention layer, the comprehensive node embedding is expected to integrate meta path based embedding as each of them preserves the proximity between nodes in corresponding semantic space.
In order to capture the personalized preference on meta paths of each node, we utilize meta path attention layer to model the personalized preference and learn the comprehensive node embedding $H$.

We first introduce a meta path preference vector $p_{i}\in R^{1\times k}$ for each node $v_{i}$ to guide the meta path attention mechanism to distinguish meta path based embedding, it is randomly initialized and jointly learned during the training process.
For meta path based embedding $h^{\pi}_{i}$ similar to the preference vector $p_{i}$, it will be assigned with large attention coefficients and contribute more in the comprehensive embedding.

To measure the similarity between the meta path preference vector and transformed meta path based embedding, we use a linear transformation with non-linear activation to transform the meta path based embedding $H^{\pi}$ into $k$-dimension space:
\begin{equation}
	{h^{\pi}_{i}}^{'}=\sigma(\mathbf{W}_{p}h^{\pi}_{i}+b_{p})
\label{mpa1}
\end{equation}  
where $\mathbf{W}_{p}\in R^{d\times k}$ is the parameter of transformation, ${h^{\pi}_{i}}^{'}$ is the transformed meta path based embedding, $b_{p}$ is the bias parameter of the transformation, $\sigma()$ is the activation function and we use Tanh here.
The meta path attention coefficients are then based on the similarity between preference vector and transformed meta path based embedding:
\begin{equation}
 	{\gamma^{\pi}_{i}}^{'}=\frac{p^{T}_{i}\cdot {h^{\pi}_{i}}^{'}}{||p_{i}||\cdot||{h^{\pi}_{i}}^{'}||}\quad
 	\gamma^{\pi}_{i}=\frac{exp({\gamma^{\pi}_{i}}^{'})}{\sum^{M}_{m=1}exp({\gamma^{m}_{i}})}
\label{mpa2}
\end{equation}
where $||\cdot||$ is the L2 normalization of vectors, $\gamma^{\pi}_{i}$ is the personalized attention coefficients on meta path $\pi$ for node $v_{i}$.

With the learned attention coefficients, the comprehensive embedding $h_{i}$ of node $v_{i}$ can be obtained by weighted combining meta path based embedding $\{h^{1}_{i},h^{2}_{i}...h^{M}_{i}\}$:
\begin{equation}
	h_{i}=\sum_{\pi =1}^{M}\gamma^{\pi}_{i}h^{\pi}_{i}
\label{mpa3}
\end{equation}
where $h_{i}$ is the comprehensive embedding of node $v_{i}$, $M$ is the number of meta paths.
To summarize, the meta path attention layer models personalized preference on meta paths by introducing the preference vector $p_{i}$ for each node $v_{i}$, the learned comprehensive embedding can naturally distinguish the meta path based embedding.

\subsection{Loss function}
In order to learn useful, predictive representations of nodes in HIN, we learn the parameters of \model\ in a task-specific semi-supervised environment.
The objective is to minimize the Cross-Entropy loss between the ground-truth and the predictions:
\begin{equation}
	L=-\sum^{|\mathcal{V}|}_{i}\sum^{L}_{l=1}Y_{il}ln(C(H_{i})_{l})
\end{equation}
where $Y_{il}\in \{0,1\}$ is ground truth of node $v_{i}$ on label $l$, $C(H_{i})_{l}\in \{0,1\}$ is predicted result of node $v_{i}$ on label $l$. 
Given partial labels of nodes, we can optimize the \model\ model with mini-batch stochastic gradient descent and back propagation algorithm. 
The overall \model\ model is described in Algorithm \ref{algo}.

\subsection{Complexity Analysis}
The overall algorithmic complexity of \model\ in single minibatch is $O(SMC_{MLP})$ where M is number of meta paths, S is the minibatch size and $C_{MLP}$ is the cost of an evaluation of the MLPs which is of the form $O(KD^{2})$ where K is the total number of layers and D is the average dimension of the layers of the MLPs in the model.
These complexities make \model\ extremely appe1aling and can be scaled to large HINs.

\begin{algorithm}
  \caption{\model\ algorithm.}
  \label{algo}
  \KwIn{HIN $G=\{\mathcal{V},\mathcal{E}\}$, meta path set$\{\pi_{1},\pi_{2},\pi_{3}...\pi_{M}\}$, transformed feature dimension $d$, preference vector dimension $k$.}
  \KwOut{Vector representation $\mathbf{H}$}
  Randomly initialize the preference vector $p_{i}\in R^{1\times k}$ for each node $v_{i}$, transformation matrix $\mathbf{W}_{f},\mathbf{W}_{C},\mathbf{W}_{T}$\;
  \While{Not converged}
  {
  \For{$v_{i}\in \mathcal{V}$}
    {
        \For{$\pi \in \{1,2,...,M\}$}
        {
            Learn path instance attention coefficients $\alpha^{\pi}$ by Equation (\ref{nal1})\;
            Learn comprehensive node embedding $\mathbf{H}$ by Equation (\ref{nal2})(\ref{nal3})\;
        }
    
    \For{$\pi \in \{1,2,...,M\}$}
    {
        Learn meta path attention coefficients $\gamma^{\pi}$ by Equation (\ref{mpa1}), (\ref{mpa2})\;
        Learn comprehensive node embedding $\mathbf{H}$ by Equation (\ref{mpa3})\;
    }
    }
    Update the parameters $p,\mathbf{W}_{f},\mathbf{W}_{C},\mathbf{W}_{p},b_{p}$;
  }
  Return node embedding $\mathbf{H}$\;
\end{algorithm}

\section{Experiment}
\label{sectionexp}
In this section, we conduct extensive experiments on real-world HIN datasets to answer the following questions:
\begin{enumerate}
    \item Can \model\ learn better HIN embedding results for data mining tasks compared with existing methods?
    \item Can \model\ model the personalized preference on meta paths and are the learned attention coefficients correct?
    \item Is \model\ sensitive to the parameters and how does the parameters affect the performance?
\end{enumerate}. 

\begin{table}[]
\caption{Dataset statistics}
\begin{tabular}{|c|c|c|c|c|c|}
\hline
       & \# Node  & \# Edge & \# Label & meta-path                                                       \\ \hline
DBLP   & 10650   & 39888   & 4       & \begin{tabular}[c]{@{}c@{}}APA,APPA,\\ APTPA,APVPA\end{tabular} \\ \hline
YELP  &  37342       & 178516        & 3         &BRURB,BRKRB                                                              \\ \hline
IMDB  &  44634      & 134643        &21          &MAM,MDM,MUM                                                                 \\ \hline
\end{tabular}
\label{dataset}
\vspace{-0.2in}
\end{table}

\subsection{Datasets}
We conduct experiments on the following real world HINs, the details of datasets are described as follows with dataset statistics summarized in Table \ref{dataset}:
\begin{enumerate}
\item \textbf{DBLP} is a bibliographic network of computer science which is frequently used in the study of heterogeneous networks. 
It contains four types of objects including paper, author, venue and topic. 
We use a subset of DBLP containing 4249 papers(P), 1909 authors(A) and 18 venues(V) from 4 areas: database, data mining, machine learning and information retrieval. 
We consider the meta path set: APA, APPA, APVPA.

\item \textbf{IMDB} is a movie rating website contains a social network of users and the rating of each user. 
We extract a heterogeneous information network with 942 users(U), 1,318 movies(M),  889 directors(D) and 41,485 actors(A). 
The type of the movie is used as the label of the movie. 
We consider the meta path set: MAM, MUM.

\item \textbf{Yelp-Restaurant} is a social media dataset, released in Yelp Dataset Challenge \footnote{\url{https://www.yelp.com/dataset/challenge}}. 
We extracted information related to the restaurant business objects of three sub-categories \cite{Li:2017:SCA:3038912.3052576} : Fast Food, Sushi Bars and American(New) Food. 
We construct a HIN of 2,614 business objects (B); 33,360 review objects (R); 1,286 user objects (U) and 82 food relevant keyword objects (K). 
We consider the meta path set BRKRB,BRURB.
\end{enumerate}

\begin{table*}[htbp]
\caption{Node classification on real world datasets.}
\vspace{-0.1in}
\label{trans}
\scalebox{0.73}{
\begin{tabular}{|c|c|c|c|c|c|c|c|c|c|c|c|c|c|c|c|c|c|c|}
\hline
Dataset                   & \multicolumn{6}{c|}{DBLP}                                     & \multicolumn{6}{c|}{YELP}                                     & \multicolumn{6}{c|}{IMDB}                                     \\ \hline
Evaluation                & \multicolumn{3}{c|}{Micro-F1} & \multicolumn{3}{c|}{Macro-F1} & \multicolumn{3}{c|}{Micro-F1} & \multicolumn{3}{c|}{Macro-F1} & \multicolumn{3}{c|}{Micro-F1} & \multicolumn{3}{c|}{Macro-F1} \\ \hline
Seed \%                   & 10\%     & 20\%     & 30\%    & 10\%     & 20\%     & 30\%    & 10\%     & 20\%     & 30\%    & 10\%     & 20\%     & 30\%    & 10\%     & 20\%     & 30\%    & 10\%     & 20\%     & 30\%    \\ \hline
Node2Vec                  & 0.841    & 0.848    & 0.858   & 0.842    & 0.849    & 0.859   & 0.858    & 0.862    & 0.867   & 0.858    & 0.867    & 0.870   & 0.421    & 0.428    & 0.435   & 0.094    & 0.102    & 0.108   \\ \hline
LINE                      & 0.845    & 0.852    & 0.852   & 0.845    & 0.853    & 0.853   & 0.859    & 0.864    & 0.871   & 0.859    & 0.867    & 0.872   & 0.427    & 0.431    & 0.440   & 0.102    & 0.112    & 0.120   \\ \hline
GraphSAGE                 & 0.853    & 0.857    & 0.866   & 0.853    & 0.858    & 0.866   & 0.883    & 0.894    & 0.899   & 0.889    & 0.898    & 0.906   & 0.458    & 0.461    & 0.468   & 0.118    & 0.119    & 0.120   \\ \hline
GCN                       & 0.851    & 0.856    & 0.863   & 0.848    & 0.852    & 0.864   & 0.883    & 0.890    & 0.892   & 0.893    & 0.900    & 0.899   & 0.468    & 0.473    & 0.476   & 0.139    & 0.143    & 0.150   \\ \hline
GAT                       & 0.841    & 0.859    & 0.869   & 0.840    & 0.859    & 0.869   & 0.889    & 0.897    & 0.899   & 0.901    & 0.907    & 0.909   & 0.475    & 0.481    & 0.486   & 0.162    & 0.170    & 0.169   \\ \hline
Metapath2Vec              & 0.766    & 0.788    & 0.799   & 0.765    & 0.789    & 0.800   & 0.860    & 0.866    & 0.870   & 0.861    & 0.867    & 0.871   & 0.427    & 0.430    & 0.434   & 0.124    & 0.129    & 0.133   \\ \hline
HIN2Vec                   & 0.853    & 0.860    & 0.862   & 0.854    & 0.861    & 0.862   & 0.869    & 0.870    & 0.871   & 0.869    & 0.876    & 0.878   & 0.437    & 0.436    & 0.437   & 0.129    & 0.127    & 0.127   \\ \hline
\model-avg & 0.842    & 0.852    & 0.858   & 0.847    & 0.854    & 0.861   & 0.883    & 0.892    & 0.899   & 0.889    & 0.895    & 0.901   & 0.467    & 0.472    & 0.486   & 0.125    & 0.133    & 0.138   \\ \hline
\model-max & 0.830    & 0.803    & 0.821   & 0.831    & 0.804    & 0.822   & 0.872    & 0.882    & 0.888   & 0.874    & 0.886    & 0.891   & 0.439    & 0.446    & 0.458   & 0.098    & 0.106    & 0.115   \\ \hline
\model     & 0.864    & 0.886    & 0.893   & 0.864    & 0.887    & 0.893   & 0.902    & 0.923    & 0.938   & 0.913    & 0.931    & 0.945   & 0.480    & 0.484    & 0.498   & 0.141    & 0.147    & 0.152   \\ \hline
\end{tabular}}
\vspace{-0.1in}
\end{table*}

\subsection{Baselines}
We evaluate \model\ against six state-of-the-art network embedding models and two \model\ variants. 
\begin{enumerate}
	\item \textbf{Node2vec/LINE} \cite{grover2016node2vec,tang2015line} Node2vec uses truncated random walks to generate node sequences and employ skip-gram model for node representation learning. 
	LINE learns node embedding by preserving both first order and second order proximity. 
	Both of two methods are widely used as the baseline of network embedding methods.
	\item \textbf{GraphSAGE} \cite{hamilton2017inductive} learns node embedding by aggregating local neighborhood features. We compare with this method to demonstrate the superior of learning neighbor attention and combine different meta paths. Here, we use the mean aggregator version of GraphSAGE to show the importance of learning weights for neighbors.
	\item \textbf{GCN} \cite{kipf2016semi} learns node embedding with graph convolutional network that designed in homogeneous network. It combines both attributes and network structure.
	\item \textbf{GAT} \cite{velickovic2017graph} is an attention based network embedding method. The attention coefficients are learned by a single-layer feedforward neural network. We compared with this method to demonstrate the superior of the proposed meta attention layer.
	\item \textbf{Metapath2Vec} \cite{dong2017metapath2vec} is one of the state-of-the-art HIN embedding methods. The meta path guided random walks are performed on HIN for context generation.
	\item \textbf{HIN2vec} \cite{fu2017hin2vec} is another state-of-the-art heterogeneous information network embedding method. It learns the embedding through a deep neural network by considering the meta path. However, it does not consider the weight of different meta paths. We compare with this method to demonstrate the superior of learning meta path attention.
	\item \textbf{\model-max/\model-avg} are variants of \model\ in which we use max pooling/mean pooling to learn the comprehensive embedding from meta path based embedding. We compare with this method to demonstrate the superior of using meta path attention layer to distinguish meta paths. 
\end{enumerate}

We implement the proposed \model\ model with Pytorch, the codes has been released in Github\footnote{\url{https://github.com/zhoushengisnoob/HAHE}}.
The model parameters are randomly initialized  with a xavier initializer and Adam optimizer is employed for optimization. 
We use the parameters learned by \model-homo as the initialization of neighbor attention layer.
We set the learning rate to 0.0005 and the batch size to 512. 
The vector dimension of all the methods is 128. 
For the compared methods, we use the code provided by authors. 
For random walk based method, the number of walks per node is set to 80, the walk length is 100, the size of negative sampling is 5. 
All the experiments are conducted on a Linux server with one NVIDIA Titan Xp GPU and 24 core Intel Xeon E5-2690 CPU.
For methods designed for homogeneous information network, to make fair comparison, we extract the homogeneous information network based on each meta path and report the best results in concatenation, max-pooling and single meta path based embedding.

\subsection{Node Classification}
Node classification has been widely used in literature to evaluate network embeddings.
The structural features are used as node attributes for GraphSAGE, GAT and GCN. 
We use Micro-F1 and Macro-F1 score as the evaluation metric for classification.
Table \ref{trans} illustrated the semi-supervised node classification results in three datasets.
Based on the results, we have the following observations:
\begin{enumerate}
    \item An overall observation is that \model\ achieves the best performance among the compared algorithms and \model\ variants in terms of all the evaluation metrics. With more labeled data for classification, most of the methods get better performances. This indicates the effectiveness of our proposed \model\ model in learning node embeddings in HIN.
    
    \item Among the methods deigned for homogeneous information network, GraphSAGE, GAT and GCN gains better performance compared with Node2Vec and LINE. This demonstrates the superior of collecting information from connected nodes. Also, GAT gains slight improvements over GraphSAGE which points out that learning weights for neighborhood can help learn better representations. 
    
    \item Compare \model\ with its two variants: \model-avg and \model-max, we observe that learning attention for aggregating meta path based embedding can achieve better performance. Mean-pooling and max-pooling are two popular pooling function for aggregating features but the importance of each meta path is ignored. The performance improvement gained by \model\ further indicates the advantage of distinguishing meta path based embedding.
\end{enumerate}

\begin{figure}[htbp]
\centering
\subfigure[\scriptsize{Metapath2Vec}]{
\centering
\epsfig{file=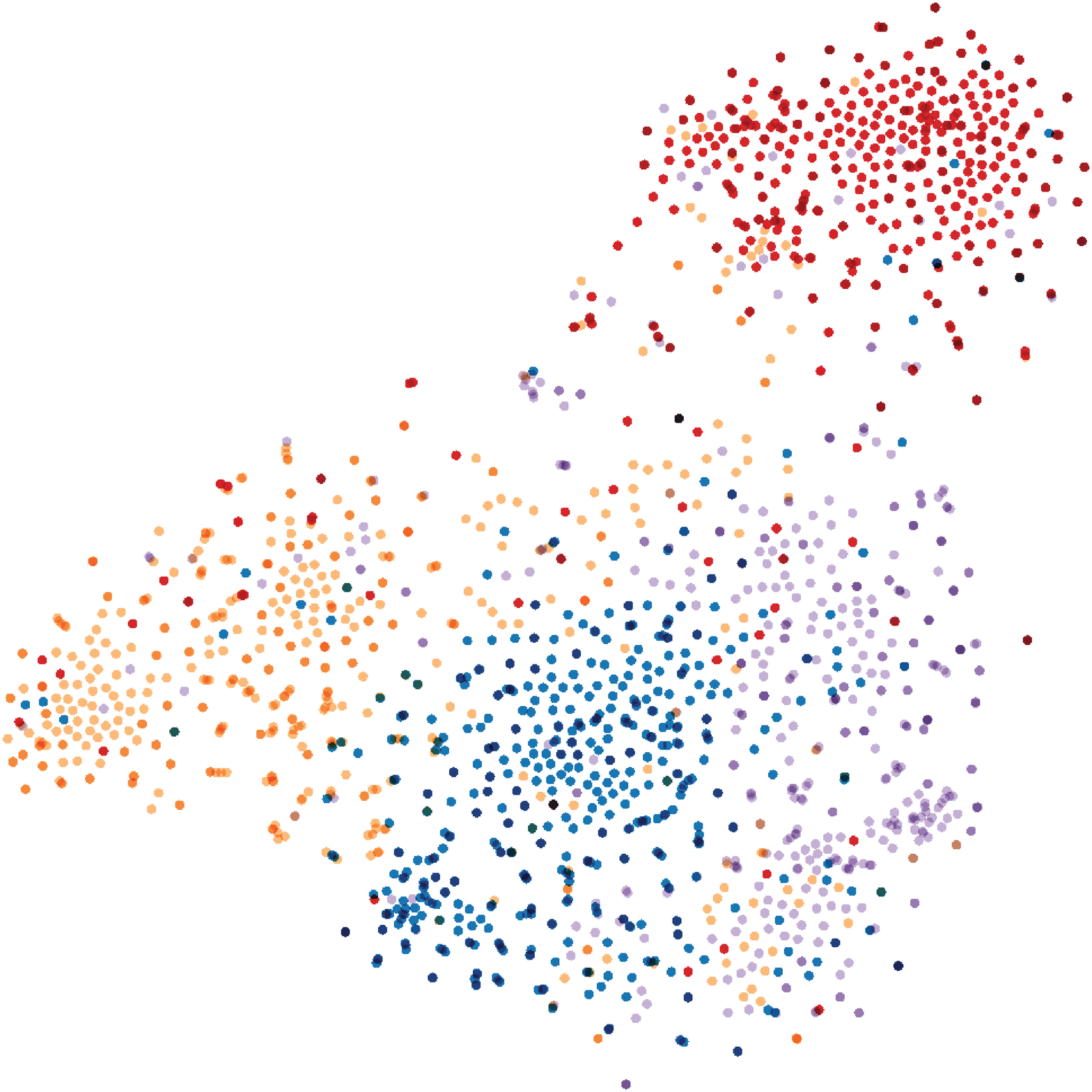,width=0.3\columnwidth}
}
\subfigure[\scriptsize{HIN2Vec}]{
\centering
\epsfig{file=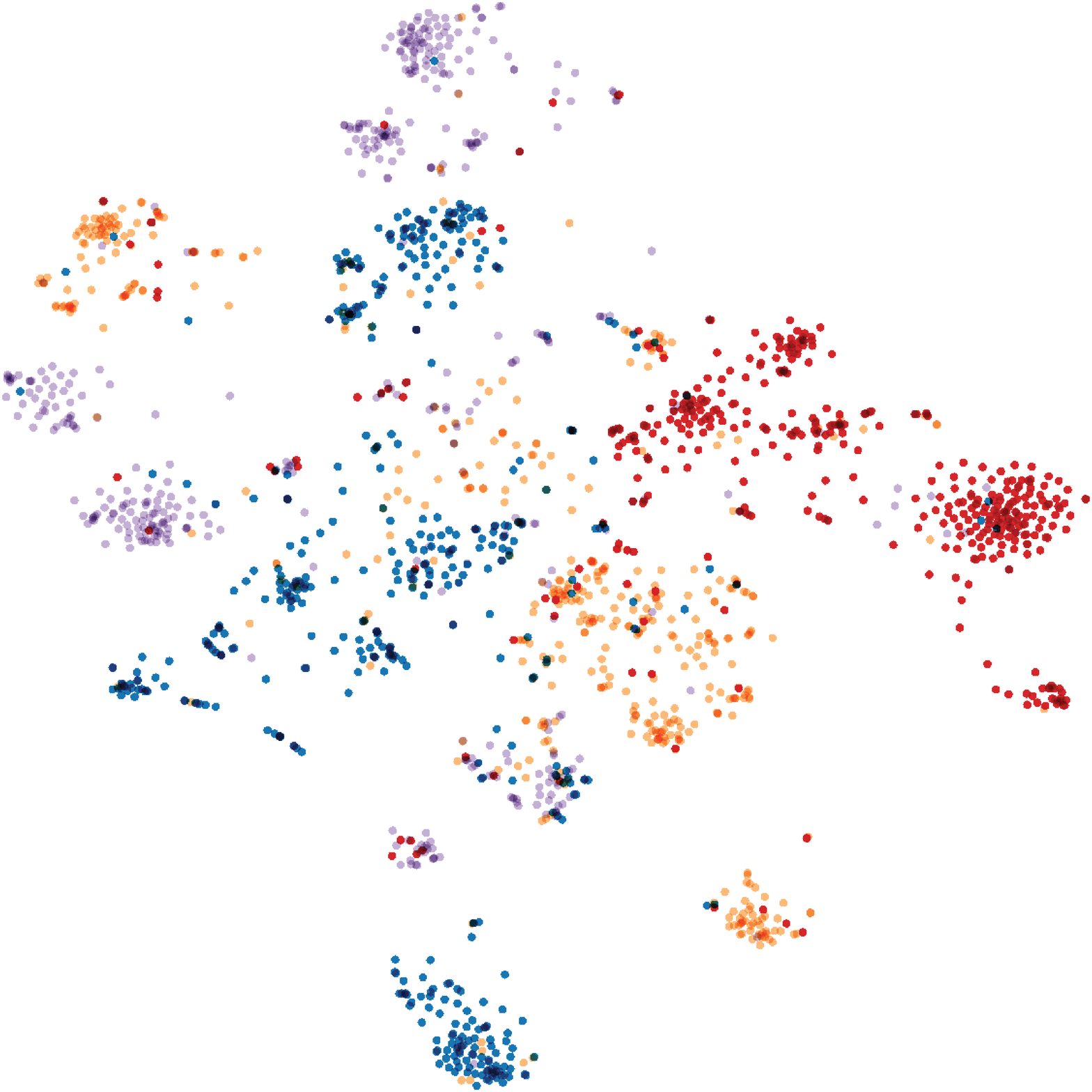,width=0.3\columnwidth}
}
\subfigure[\scriptsize{HAHE}]{
\centering
\epsfig{file=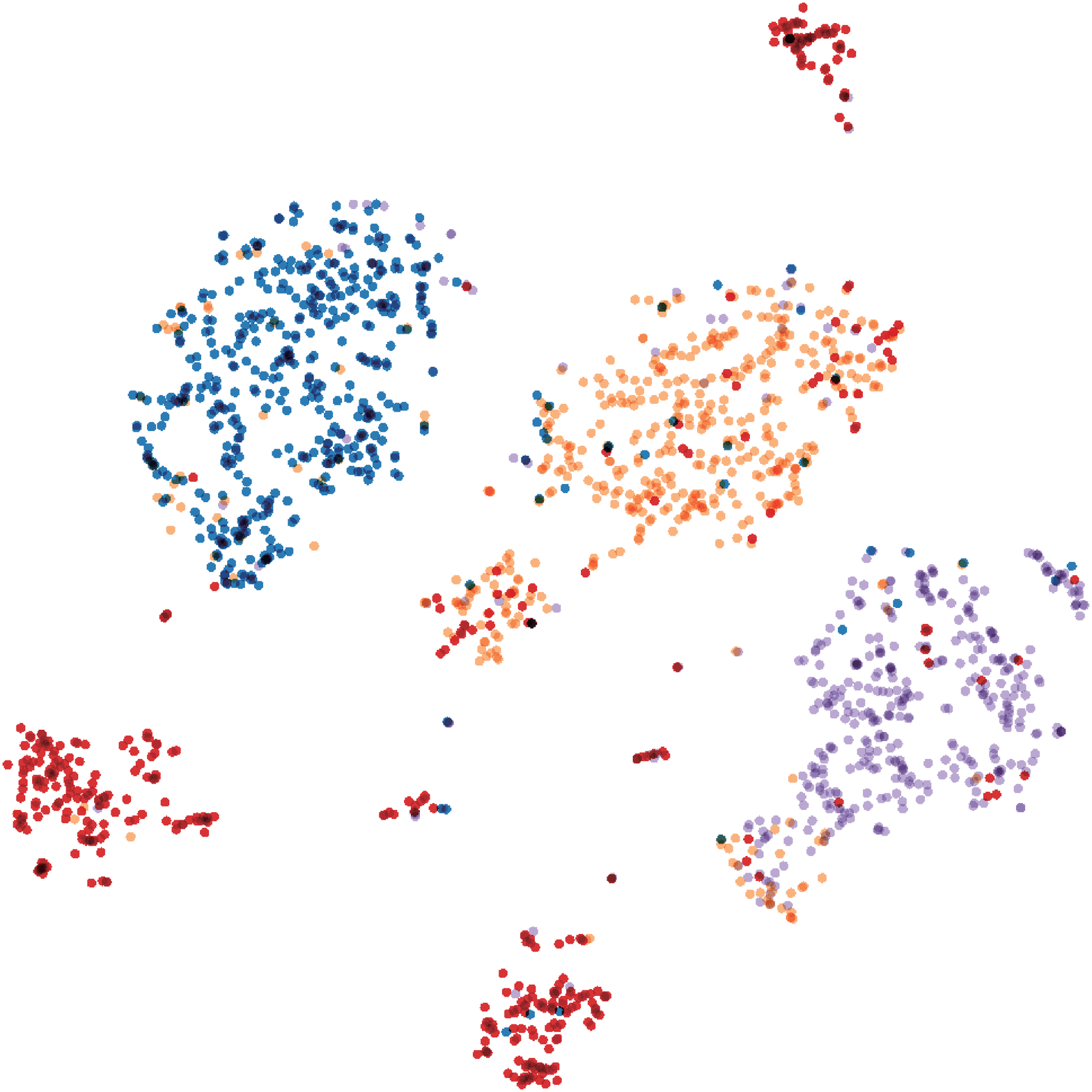,width=0.3\columnwidth}
}
\vspace{-0.1in}
\caption{Network Visualization results on DBLP dataset. Nodes are mapped into the 2-D space using the T-SNE package with learned embeddings. Color indicates the class label. Best viewed on screen.}
\label{vis}
\vspace{-0.1in}
\end{figure}

\subsection{Network Visualization}
Network visualization is another popular application of network embedding which supports tasks such as data exploration and understanding. 
Following the experimental setting of existing works \cite{zhou2018prre}, we first learn low dimensional representation for each node and then map them into the 2-D space with t-SNE\cite{maaten2008visualizing}. Figure \ref{vis} illustrates the network visualization results on DBLP dataset. Each dot denotes a node and each color denotes label of a class. 
A good embedding method is expected to make nodes with same label close to each other while far for nodes with different labels.
As observed in Fig \ref{vis}, the state-of-the-art baseline methods Metapath2Vec and HIN2Vec do not separate the nodes as good as \model.  
The visualization results of \model\ are quite clear since most of nodes with same label (color) are close to each other and nodes with different labels(colors) are far from each other. This further verifies the effectiveness of the proposed \model\ method.

\subsection{Analysis of attention coefficients}
\begin{figure}[htb!]
\centering
\subfigure[\scriptsize{DBLP}]{
\centering
\epsfig{file=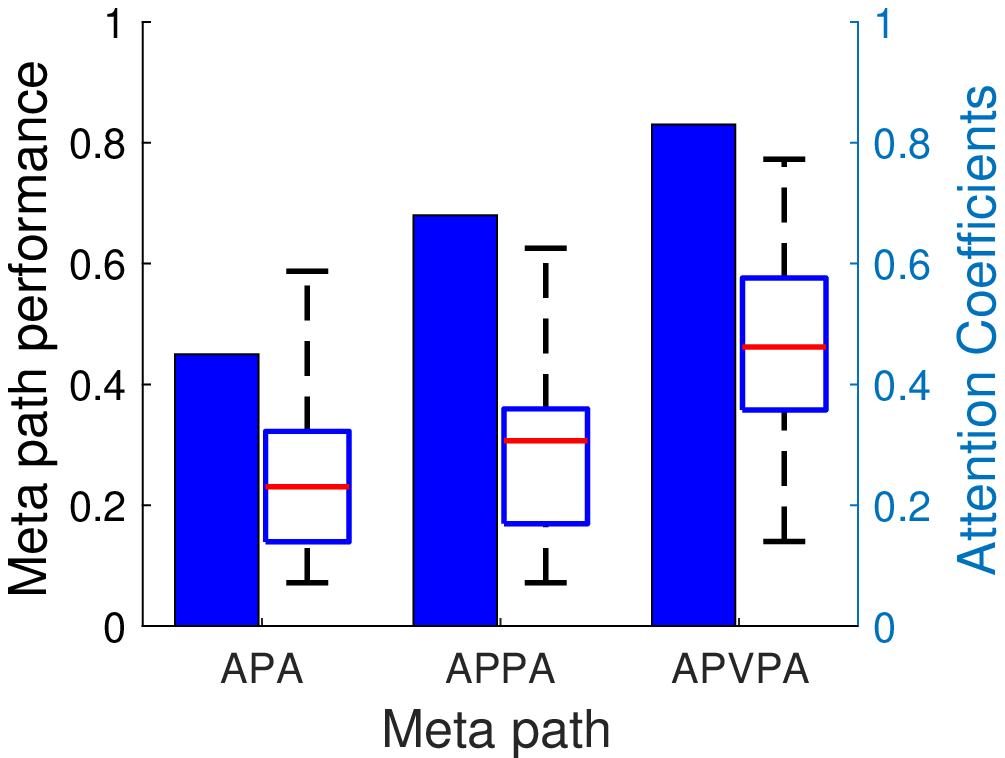,width=0.364\columnwidth}
}
\subfigure[\scriptsize{YELP}]{
\centering
\epsfig{file=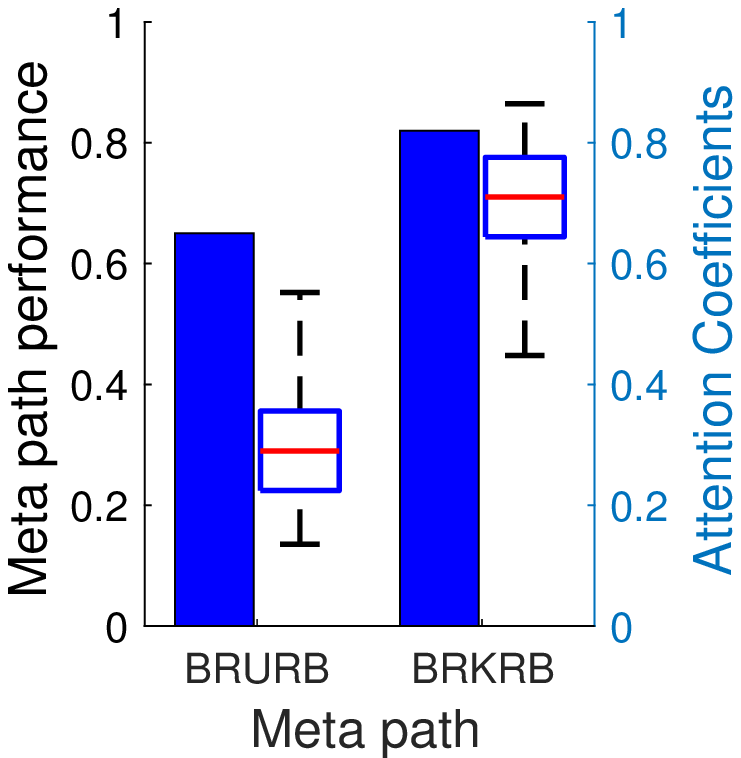,width=0.268\columnwidth}
}
\subfigure[\scriptsize{IMDB}]{
\centering
\epsfig{file=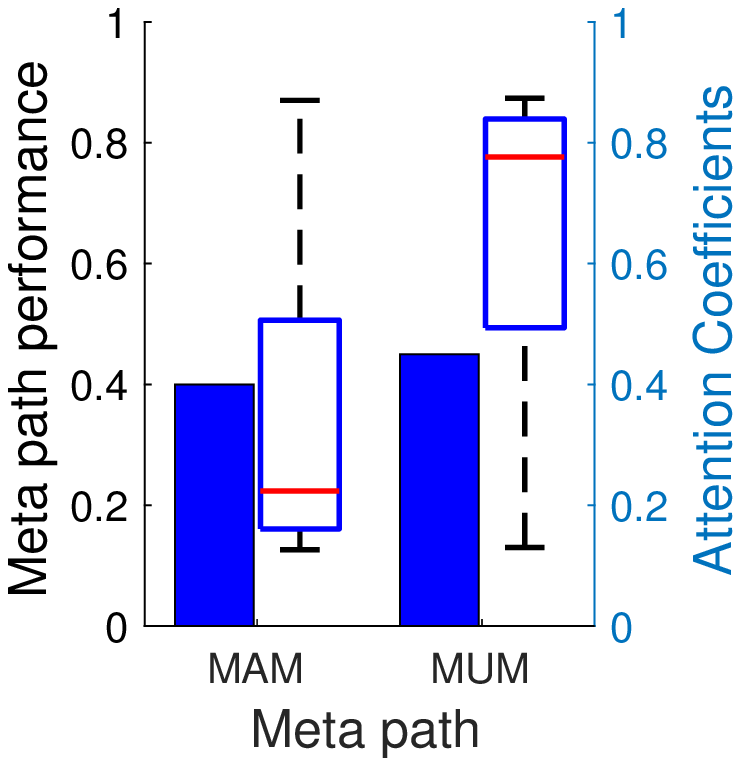,width=0.268\columnwidth}
}
\vspace{-0.1in}
\caption{Comparison between performance of meta path and learned attention coefficients. The bar plot(left) denotes the meta path performance and the box plot(right) denotes the learned attention coefficients.}
\label{atten}
\vspace{-0.1in}
\end{figure}

In \model\ model, we use meta path attention layer to model the personalized preference on the meta paths. 
To evaluate whether the learned attention coefficients could reflect the personalized preference on meta paths, we compared the learned attention coefficients with the performance of meta paths.
The performance of each meta path can be represented by single meta path based embedding which is the output of the neighbor attention layer in \model. We directly fed the meta path based embedding into node classification task and Figure \ref{atten} illustrates the comparison between performance of meta path and learned attention coefficients. 
Based on the results, we have the following observations:
\begin{enumerate}
	\item The basic observation is that there is a positive correlation between the performance of meta path and learned attention coefficients. Meta path with better performance is assigned with larger average attention coefficients. This proves that the learned attention coefficients can properly reflect the performance of meta paths. As a result, users can have deeper insight of the dataset.
	\item Another interesting observation is that the box plot \footnote{\url{https://en.wikipedia.org/wiki/Box_plot}} of each meta path shows that nodes have personalized preference on meta paths. Although the trend of attention coefficients are same as meta path instances, nodes has different attention coefficients on each meta path. Such personalized preference is captured by \model\ to learn more accurate embedding.
	\end{enumerate}
 
\subsection{Parameter Analysis}
In this subsection, we investigate the parameter sensitivity of \model.
More specifically,  we evaluate how different numbers of the node embedding and preference vector dimension can affect the results of node classification. 
Following the previous experiment settings\cite{wang2016structural}, we only change the corresponding dimension and report the results.

\begin{figure}[htb!]
\centering
\subfigure[\scriptsize{Embedding dimension $d$}]{
\centering
\epsfig{file=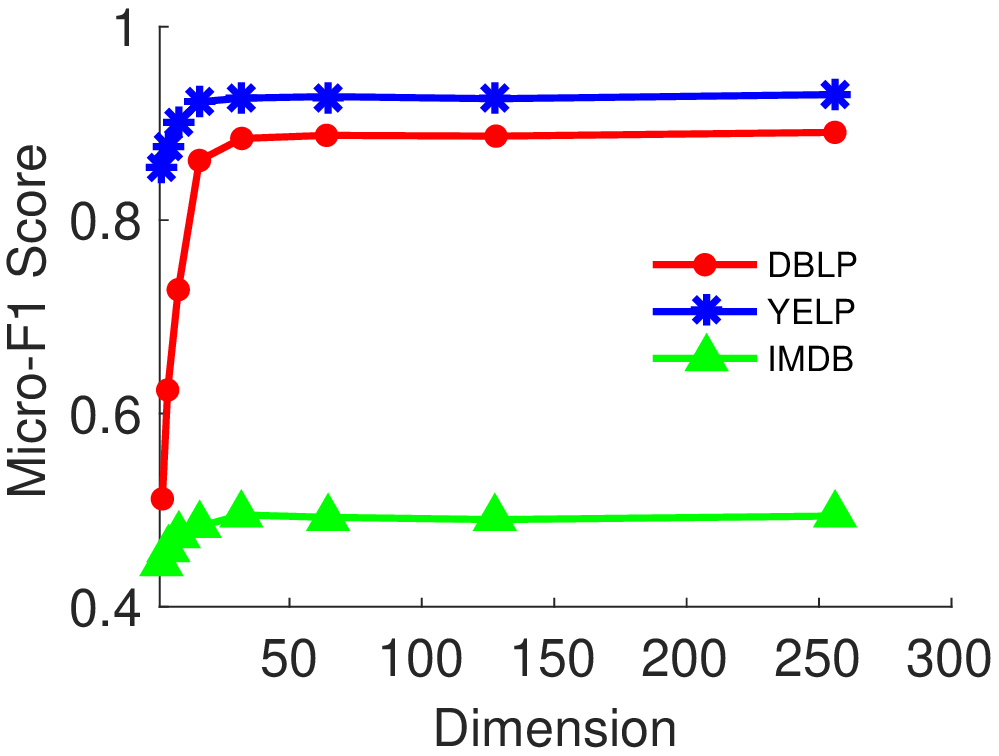,width=0.45\columnwidth}
}
\subfigure[\scriptsize{Preference vector dimension $k$}]{
\centering
\epsfig{file=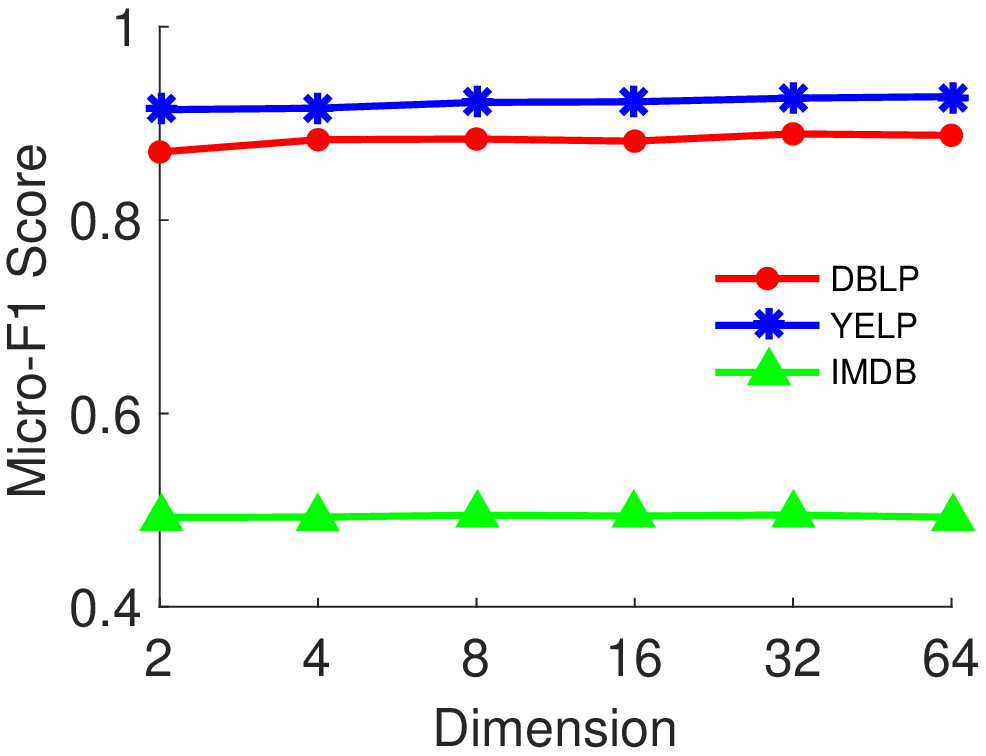,width=0.45\columnwidth}
}
\vspace{-0.1in}
\caption{Parameter sensitivity analysis w.r.t dimension of node embedding $h$ and preference vector $p$.}
\label{hidden}
\vspace{-0.1in}
\end{figure}

Figure \ref{hidden} illustrates the result of Micro-F1 score w.r.t dimension of embedding and preference vector.
In Figure \ref{hidden}(a), we can see that the dimension slightly affect the classification performance, when the number of dimensions is larger than 50, the Micro-F1 score have minor changes(within 1\%) which shows that \model\ is not very sensitive to the dimension of context vector.
In Figure \ref{hidden}(b), we have similar observation that \model\ is not sensitive to the dimension of preference vector on three datasets.

\section{Conclusion}
\label{sectionconclu}
In this paper, we have proposed the \model model for HIN embedding in which a hierarchical attention mechanism is proposed to model the personalized preference on both meta paths and path instances.
Experimental results of node classification, network visualization on real-world HIN datasets demonstrate the superior performance of \model compared with several state-of-the-art methods.

This paper suggests several potential future directions of research. First, the recommender system can be viewed as a HIN with rich attribute information, the problem of recommending items or friends to users can be based on the embedding of items and users. Another possible direction is to take the attributes of nodes into consideration since in real-world datasets, nodes are often associated with rich information. Finally, learning embedding for all types of nodes in HINs is also an interesting problem to solve.

\bibliographystyle{ACM-Reference-Format}
\bibliography{reference}


\end{document}